\documentclass[9pt, final, technote]{IEEEtran}
\def\tc#1{#1}
\def\oc#1{}
\usepackage[latin1]{inputenc}
\usepackage{algorithm}
\usepackage{algpseudocode}
\usepackage{tikz}
\usepackage{pgfbaselayers}
\usepackage{colortbl}

\usepackage{bm} 
\usepackage{amsmath}
\usepackage{amsfonts}
\usepackage{amssymb}
\usepackage{mathrsfs}
\usepackage{dsfont}
\usepackage{mathtools}
\mathtoolsset{showonlyrefs}

\usepackage{algorithm}
\usepackage{algpseudocode}

\usepackage[squaren,thinspace]{SIunits}

\usepackage{theorem}
\theoremheaderfont{\bfseries}
\theoremstyle{plain}
{\theorembodyfont{\itshape} \newtheorem{theorem}     {Theorem}}
{\theorembodyfont{\itshape} }
{\theorembodyfont{\itshape}\newtheorem{corollary}   {Corollary}}
{\theorembodyfont{\sffamily\small}}
\newenvironment{Proof}
{\noindent{\scshape Proof.}}
{\hspace*{\fill}$\square$}

\newcommand{\rv}[1]{\ensuremath{{\boldsymbol{#1}}}}
\renewcommand{\vec}[1]{\ensuremath{{\underline{#1}}}}
\newcommand{\rvec}[1]{\ensuremath{{\boldsymbol{\underline{#1}}}}}
\newcommand{\mat}[1]{{\ensuremath{{\mathbf{#1}}}}}

\newcommand{\argmin}[1]{\mathop{\mathrm{arg}\; \min_{#1}}}

\DeclareMathOperator{\diag}{diag}

\def\T{^\mathrm{T}} 

\def\atan2{\operatorname{atan2}}

\def\({\left(}
\def\){\right)}
\def\[{\left[}
\def\]{\right]}

\newcommand{\NewN}{\ensuremath{\mathds{N}}}

\newcommand{\NewR}{\ensuremath{\mathds{R}}}

\newcommand{\Gauss}{{\mathcal{N}}}

\newcommand{\Second}[2][]{\unit{#2}#1\second}

\def\x{\rv x}
\def\y{\rv y}

\def\vx{\vec x}

\def\rvv{\rvec v}
\def\rvw{\rvec w}
\def\rvx{\rvec x}

\def\rvz{\rvec z}

\def\A{\mat A}

\def\C{\mat C}

\def\H{\mat H}
\def\I{\mat I}

\def\M{\mat M}

\def\V{\mat V}
\def\W{\mat W}

\def\hvx{\hat{\vec x}}

\def\hvz{\hat{\vec z}}

\def\SS{\mathcal S}
\def\SU{\mathcal U}

\def\BB{branch-and-bound}
\def\IBP{information-based pruning}
\def\KF{Kalman filter}
\def\PSD{positive semi-definite}
\def\RE{Riccati equation}
\def\SIE{sensor information ellipsoid}
\def\SIM{sensor information matrix}
\def\SIMs{sensor information matrices}

\algrenewcommand{\algorithmiccomment}[1]{\hfill// #1}

\def\GRE{{GRE}}
\def\CVX{{CVX}}
\def\ZB{{ZB}}
\def\COV{{COV}}
\def\SIP{{SIM}}
\def\IBP{{IBP}}

\def\Sec#1{Section~\ref{#1}}

\def\Fig#1{Fig.~\ref{#1}}

\def\Theo#1{Theorem~\ref{#1}}

\def\Alg#1{Algorithm~\ref{#1}}

\usetikzlibrary{shapes}

\tikzstyle{hn}=[draw, rectangle, inner sep=5pt, fill=black!10!white]				
\tikzstyle{hng}=[hn, fill=\gcolor!20]

\tikzstyle{hnlines}=[scale=.8]								
\tikzstyle{hnl}=[hnlines, anchor=base east]		
\tikzstyle{hnm}=[hnlines, anchor=base north] 		
\tikzstyle{hnr}=[hnlines, anchor=base west] 		

\tikzstyle{hl}=[ draw, -latex]			
\tikzstyle{hly}=[ draw, -latex]

\tikzstyle{gnode}=[]
\tikzstyle{gnl}=[gnode, anchor=mid east]
\tikzstyle{gnr}=[gnode, anchor=mid west]

\newcounter{N}

\begin{document}

\title{\LARGE\bf{Optimal Multi-Step Sensor Scheduling for Linear Dynamical Systems}}
\title{\LARGE\bf{Optimal Pruning for Multi-Step \tc{\\} Sensor Scheduling}}

\author{%
Marco F. Huber,~\IEEEmembership{Member,~IEEE}%
\thanks{M.~F.~Huber is with the AGT Group (R\&D) GmbH, Darmstadt, Germany. Email: {\tt marco.huber@ieee.org}}
}

\maketitle


\begin{abstract}
In the considered linear Gaussian sensor scheduling problem, only one sensor out of a set of sensors performs a measurement. To minimize the estimation error over multiple time steps in a computationally tractable fashion, the so-called information-based pruning algorithm is proposed.
It utilizes the information matrices of the sensors and the monotonicity of the Riccati equation. This allows ordering sensors according to their information contribution and excluding many of them from scheduling. Additionally, a tight lower  is calculated for branch-and-bound search, which further improves the pruning performance.
\end{abstract}



\section{Introduction}
\label{sec:introduction}
In sensor systems consisting of multiple sensors, e.g., in sensor networks, most information about an observed process is obtained if each sensor performs measurements at each time instant. 
But due to constrained communication bandwidth, simultaneously sending all acquired data may not be possible. 
Furthermore, to save energy, sensors should be switched off if their measurements are not required.
To increase the operational lifetime of such sensor systems, the measurement rate should be as low as possible, which on the other hand increases the estimation error. \emph{Sensor scheduling} constitutes a promising solution to this trade-off. Here, only one sensor per time step is activated in order to minimize a function of the estimation error over a finite time horizon.

In this paper, the sensor scheduling problem for discrete-time linear Gaussian systems is studied. 
Determining the optimal \emph{sensor schedule}, i.e., the sequence of active sensors over the considered time horizon that minimizes the estimation error, requires traversing a search tree as shown in \cite{Meier_TAC1967}. 
The depth and branching factor of the tree are equal to the time horizon length and the number of sensors, respectively. Each path from the root to a leaf node of the tree corresponds to one possible sensor schedule. 
The tree-search corresponds to solving a binary integer program and thus, sensor scheduling is NP-hard.
Especially for long time horizons and/or a large number of sensors, an exhaustive tree-search can be avoided by employing pruning techniques.
Existing pruning techniques can be classified into suboptimal and optimal methods. By employing optimal pruning \cite{Logothetis1999, Fusion08_Huber_PLSS, Huber_CIP2010}, deleting the optimal sensor schedule is impossible, but many complete sensor schedules may still be evaluated. Drastic savings in computational demand arise from suboptimal pruning \cite{Alriksson_IFAC2005, Gupta_ICASSP2004}, 
but ending up with the optimal schedule is no longer guaranteed.
The sub-optimal methods in \cite{Huber_CIP2010, Joshi_TSP2009, Mo_Automatica2011} utilize convex relaxation and optimization instead of traversing the tree. In \cite{He_CDC2004, Chen_IFAC2008}, the sensor scheduling problem is formulated as partially observable Markov decision process and solved approximately by means of discretizing the belief space.

In the following, an optimal pruning technique named \emph{information-based pruning} (IBP) is introduced aiming at pruning complete sub-trees as early as possible. To achieve this, two contributions are made: First, a partial ordering of sensors based on the so-called \SIM~(see \Sec{sec:ordering}). This allows excluding sensors from tree search merely on the basis of the measurement matrix and noise covariance matrix. Second, with the \SIMs~of the remaining sensors a so-called \emph{bounding sensor} is calculated, which provides a lower bound to the estimation error (see \Sec{sec:bounding}). Both contributions are combined in a \BB~algorithm for efficiently traversing the tree for the optimal sensor schedule.

\section{Problem Formulation}
\label{sec:problem}
%
The dynamics of the observed process are given by the linear model
\begin{equation}
	\label{eq:problem_system}
	\rvx_{k+1} = \A_k\cdot \rvx_k + \rvw_k~,
\end{equation}
where $k = 0, 1, \ldots, N-1$ is the discrete time index and $N$ is the time horizon length.\footnote{Random variables are denoted by lowercase bold-face, vectors by underlined, and matrices by uppercase bold-face letters.}  
A finite set $\mathcal S$ of sensors is considered
, where measurement $\rvz_k^i$ from sensor $i \in \mathcal S = \{1, \ldots, S\}$ is related to the latent system state $\rvx_k\in \NewR^{n_x}$ via the linear measurement model
\begin{equation}
	\label{eq:problem_sensor}
	\rvz_k^i = \H_k^i \cdot \rvx_k + \rvv_k^i~.
\end{equation}

Both $\A_k$ and $\H_k^i$ are time-variant matrices. The noise terms $\rvw_k$ and $\rvv_k^i$ are zero-mean white Gaussian with covariance matrices $\C_k^w$ and $\C_k^{v,i}$, respectively, with $\C_k^{v,i}$ being positive definite. A measurement value $\hvz_k^i$ of sensor $i \in \mathcal S$ is a realization of $\rvz_k^i$. 
The initial system state $\rvx_0 \sim \Gauss(\vx_0; \hvx_0, \C_0^x)$ at time step $k=0$ is known and assumed to be Gaussian with mean vector $\hvx_0$ and covariance matrix~$\C_0^x$. 
A sensor schedule is indicated by $u = \(u_{0}, u_1, \ldots, u_{N-1}\)$, where an element $u_k$ encodes the index of the sensor to be scheduled for measurement at time step $k$, i.e., if sensor $i$ is scheduled at time step $k$ then $u_k = i$.

For linear Gaussian systems, the \KF~is the optimal estimator of $\rvx_k$ in a minimum mean-square sense. For a given sensor schedule $u$, the state covariance matrix $\C_k^x$ evolves according to the Riccati equation (partly in information form)
\tc{
\begin{multline}
\label{eq:riccati}
	\C_{k+1}^x(u) 
	= r_{u_k}(\C_k^x(u))
	\coloneqq \C_k^w\ + \\ \A_k \Bigl(\bigl(\C_k^x(u)\bigr)^{-1} + \underbrace{\(\H_k^{u_k}\)\T \(\C_k^{v,u_k}\)^{-1}\H_k^{u_k}}_{\coloneqq \M_k^{u_k}}\Bigr)^{-1}\A_k\T~,
\end{multline}
}
\oc{
\begin{equation}
\label{eq:riccati}
	\C_{k+1}^x(u) 
	= r_{u_k}(\C_k^x(u))
	\coloneqq \C_k^w + \A_k \Bigl(\bigl(\C_k^x(u)\bigr)^{-1} + \underbrace{\(\H_k^{u_k}\)\T \(\C_k^{v,u_k}\)^{-1}\H_k^{u_k}}_{\coloneqq\, \M_k^{u_k}}\Bigr)^{-1}\A_k\T~,
\end{equation}
}
where the function $r_i(\C_k^x)$ propagates the state covariance $\C_k^x$ to the next time step given the measurement of sensor $i$. 
Further, $\M_k^{i}$ is the symmetric and \PSD~\emph{sensor information matrix}. It subsumes the information contribution of sensor  $i$ at time step $k$.


The aim of multi-step sensor scheduling is to determine the sensors for the time steps $k=0, 1,\ldots,N-1$ such that the uncertainty of the state estimate and thus, the estimation error is minimized. 
For this purpose, the optimal sensor schedule $u^* = \(u_{0}^*, \ldots, u_{N-1}^*\)$ results from solving
\oc{%
\begin{equation}
\label{eq:optimal_schedule}
	u^* = \argmin{u}\{J(u)\} \ \text{ with } \ J(u) = \sum_{k=1}^{N} g_k(\W_k\C_k^x(u)\W_k\T)~,
\end{equation}
}
\tc{
\begin{align}
\label{eq:optimal_schedule}
	u^* &= \argmin{u}\{J(u)\} \\
	\shortintertext{with}
	J(u) &= \sum_{k=1}^{N} g_k(\W_k\C_k^x(u)\W_k\T)~,
\end{align}
}
where $\W_k$ is a positive semi-definite weighting matrix. By means of $\W_k$, an adaptation to specific scheduling objectives is possible. For instance, with $\W_k = \I$ for $k=1,\ldots,N$ the focus is on minimizing the average estimation error, while for $\W_k = \mat 0$ for $k=1,\ldots,N-1$ and $\W_N = \I$ merely the terminal estimation error is minimized. 
The cost function $g_k$ can be 
the trace, determinant, or the maximum eigenvalue of the weighted state covariance matrix. 
Thus, $g_k$ maps the weighted estimation error or uncertainty subsumed in $\W_k\C_k^x(u)\W_k\T$\,, $k=1,\ldots,N$ to a scalar value. $J(u)$ is the \emph{total cost} of the schedule $u$. 

\begin{figure}[tb]
\centering
\begin{tikzpicture}[level distance=2cm]
  \tikzstyle{level 1}+=[sibling distance=4.5cm]
  \tikzstyle{level 2}+=[sibling distance=2cm]

  \tikzstyle{hng}+=[minimum size=.3cm, rounded corners=3pt]
  \tikzstyle{hnRoot}+=[minimum size=0.3cm]
      
	\path[hl] node[] (root) {$\C_{0}^x$}
		child {node[] (g1) {$\C_1^x(1)$} 
			child {node[] (g11) {$\C_2^x(1,1)$}
				edge from parent node[hnl] {$u_{1}=1$}
			}
			child {node[] (g12) {$\C_2^x(1,2)$}
				edge from parent node[hnr] {$u_{1}=2$}
			}
			edge from parent node[hnl] {$u_{0}=1$}
		}
		child {node[] (g2) {$\C_1^x(2)$}
			child {node[] (g21) {$\C_2^x(2,1)$}
				edge from parent node[hnl] {$u_{1}=1$}
			}
			child {node[] (g22) {$\C_2^x(2,2)$}
				edge from parent node[hnr] {$u_{1}=2$}
			}
			edge from parent node[hnr] {$u_{0}=2$}
		}
	;
	
		
	\end{tikzpicture}  		  
  \caption{Search tree for time horizon length $N=2$ and $S=2$ sensors with root node $\C_{0}^x$.}
  \label{fig:tree}
\end{figure}
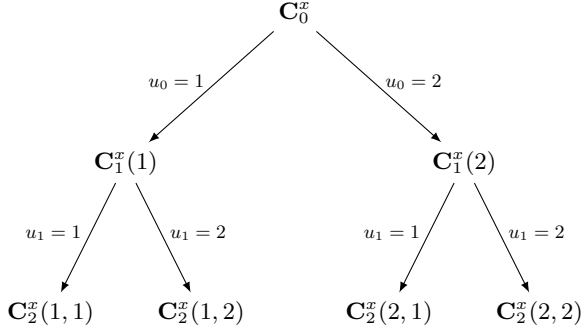

Solving the optimization problem in \eqref{eq:optimal_schedule} corresponds to finding the path with lowest cost from the root to a leaf node in a tree like the one depicted in \Fig{fig:tree}.  Obviously, enumerating all possible paths is computationally infeasible for many sensors and/or a long time horizon.

\section{Partial Ordering of Sensors}
\label{sec:ordering}
The first step of the proposed IBP is based on the monotonicity property of the Riccati equation. 

\begin{theorem}
\label{theo:properties}
For any $i \in \SS$, positive semi-definite matrix $\W_k$, and covariance matrices $\C_k^x$, $\tilde \C_k^x$ with $\C_k^x \succeq \tilde \C_k^x$, i.e., $\C_k^x - \tilde \C_k^x$ is \PSD, it holds:
\begin{description}
	\item[(Monotonicity)] \hspace{31mm} $r_i(\C_k^x) \succeq r_i(\tilde \C_k^x)$\ ,
	\item[(Cost ordering)] \hspace{31mm} $g_k(\C_k^x) \ge g_k(\tilde \C_k^x)$\ ,
	\item[(Symmetric weighting)] \hspace{31mm} $\W_k\C_k^x\W_k\T \succeq \W_k\tilde \C_k^x\W_k\T$\ .
\end{description}
%
\end{theorem}

For a proof of the monotonicity see Lemma 2 in \cite{Gupta_Automatica2006} and for the cost ordering property see \cite{Abadir_2005}. The monotonicity implies that the \PSD~ordering between covariance matrices will not change by applying the \RE. 
This ordering is also not affected by multiplying the covariance matrices with the positive semi-definite weighting matrices $\W_k$ in~\eqref{eq:optimal_schedule}. 

Furthermore, the ordering property automatically implies an ordering of the (scalar) cost functions. But pruning nodes merely on the basis of the scalar costs $g_k$ without considering the ordering of the covariance matrices automatically leads to suboptimal pruning. This follows from the fact that the cost ordering is only a necessary but no sufficient condition for the \PSD~ordering between $\C_k^x$ and $\tilde \C_k^x$. Hence, at each node of the search tree, the ordering of covariance matrices needs to be determined explicitly for the next time step in order to ensure optimal pruning. 
On this account, the optimal pruning procedure proposed in \cite{Logothetis1999} evaluates the \RE~for each sensor, which is computationally demanding for large state spaces or high dimensional measurements. The proposed pruning technique instead allows determining the order of covariance matrices without evaluating the \RE~multiple times.

\tc{%
\begin{algorithm*}[tb]
\caption{IBP branch-and-bound algorithm. The algorithm is initialized with $J_\mathrm{min} = \infty$.}
\label{alg:bb}
\begin{algorithmic}[1]
\State For a given sensor schedule $(u_0,\ldots,u_{k-1})$ do:
\If{leaf node, i.e., $k=N$}
	\State $J_\mathrm{min} \leftarrow J(u_0,\ldots,u_{N-1})$ 
\Else
	\State $\SU \leftarrow $ child($u_{k-1}$) \algorithmiccomment{children of sensor $u_{k-1}$}
	\State $\SU' \leftarrow $ prune($\SU$) \algorithmiccomment{use ordering of sensor information matrices}
	\State bound($\SU'$) \algorithmiccomment{calculate lower bounds for all sensors in $\SU'$}
	\State $\SU' \leftarrow $ sort($\SU'$) \algorithmiccomment{sort in ascending order based on lower bounds}
	\State $\SU'' \leftarrow $ prune($\SU'$) \algorithmiccomment{use lower bounds and $J_\mathrm{min}$}
	\State expand($\SU''$) \algorithmiccomment{call Algorithm 1 for remaining sensors $u_{k}\in\SU''$}
\EndIf
\end{algorithmic}
\end{algorithm*}
}

\begin{theorem}[Order of Sensor Information Matrices]
	\label{theo:sensorinfo}
	\hspace{1mm}\\Given the covariance matrix $\C_k^x$ and the sensor information matrices $\M_k^{i}$ and $\M_k^{j}$ for two sensors $i, j \in \SS$ such that
	\begin{equation}
		\label{eq:sim_ordering}
		\M_k^{i} \succeq \M_k^{j}~,
	\end{equation}
	then $r_i(\C_{k}^x) \preceq r_j(\C_{k}^x)$\,.
\end{theorem}
\begin{Proof}
	Adding $\(\C_k^x\)^{-1}$ to both sides of \eqref{eq:sim_ordering} and inverting both sides
	yields 
	%
	$\bigl(\(\C_k^x\)^{-1} + \M_k^{i}\bigr)^{-1} \preceq \bigl( \(\C_k^x\)^{-1} + \M_k^{j}\bigr)^{-1}.$
	%
	Both sides are now multiplied from right with $\A_k\T$ and from left with $\A_k$. Adding $\C_k^w$ results in	$r_i(\C_{k}^x) \preceq r_j(\C_{k}^x)$\,.
	%
\end{Proof}

Thus, for checking the \PSD~ordering of covariance matrices, merely the ordering~\eqref{eq:sim_ordering} of sensor information matrices for two sensors $i$ and $j$ needs to be determined. Selecting sensor $i$ will then provide a smaller covariance matrix at time step $k+1$. Due to the monotonicity of the Riccati equation, this even holds for arbitrary sensor schedules from time step $k+1$ on.

\begin{corollary}[Pruning via Sensor Information Matrices]
	\label{cor:infobased_pruning}
	Suppose that the sensor schedule 
	\tc{$u' = (u_0, u_1, \ldots, u_{k-1})$} 
	\oc{$u' = $\linebreak $(u_0, u_1, \ldots, u_{k-1})$} 
	up to time step $k-1$ results in the covariance matrix $\C_{k}^x(u')$\ . If for two sensors $u_k$ and $\tilde u_{k}$ the order 
	\tc{
	\begin{align}
		\M_k^{u_k} \preceq \M_n^{\tilde u_k}
	\end{align}
	}
	\oc{
	$\M_k^{u_k} \preceq \M_n^{\tilde u_k}$ 
	}
	holds, then $\forall\, u'' = (u_{k+1}, \ldots, u_{N-1})$ there exists a schedule $\tilde u'' = (\tilde u_{k+1}, \ldots, \tilde u_{N-1})$ such that $J\(u\) \ge J\(\tilde u\)$,	where $u = (u', u_k, u'')$ and $\tilde u = \(u', \tilde u_k, \tilde u''\)$\,.
\end{corollary}
\begin{Proof}
	Due to \Theo{theo:sensorinfo}, at least the sensor schedule $\tilde u'' = u''$ yields $\C_{k}^x(u) \succeq	\tilde \C_{k}^x(\tilde u)$ for all $k$. The cost ordering according to \Theo{theo:properties} then leads to $J(u) \ge J(\tilde u)$\,.
\end{Proof}

Without evaluating the \RE~at time step $k$ and only by comparing the sensor information matrices of sensor $u_k$ and $\tilde u_k$, it can be decided that only sensor $\tilde u_k$ needs to be considered for determining the optimal sensor schedule, while the covariance matrices $\C_{k+1}^x(u), \ldots, \C_N^x(u)$ need not to be determined for any sequence $u''$. Hence, the complete sub-tree of sensor $u_k$ can be pruned.

\subsection{Order of Sensor Information Matrices}
\label{sec:olf_pruning_order}
When comparing sensor information matrices (or equivalently covariance matrices), the difference is in some cases indefinite, i.e., it is not determinable, if one information matrix is ``larger" than the other. This is due to the fact that the order relation~$\preceq$ of positive semi-definite matrices leads to partial orders. Thus, in general, not all sensors can be pruned at a specific time step.

Scalar systems, i.e., the dimension of the state $\x$ is one, form an exception. Here, the partial order becomes a total order. Per time step, it is now possible to prune all sensors except of one, which is equivalent to selecting the sensor that minimizes the state variance at each time step. This greedy strategy automatically leads to the optimal sensor schedule. 
Similar results can be found in \cite{Howard_Fusion2004} for the case of scheduling multiple scalar systems to a single sensor.

\subsection{Branch-and-Bound}
\label{sec:olf_pruning_bb}
Even if pruning based on \SIMs~reduces the number of possible sensor schedules, the remaining number of nodes in the search tree may still be large due to the partial order. To further prune the tree, the proposed technique is combined with branch-and-bound search.
Branch-and-bound is a common search technique for classical decision problems like traveling-salesman or knapsack. The basic idea is to assign a lower bound of the achievable total cost value to any visited node. By means of these bounds, the tree is further expanded (branched), whereas nodes with a smaller lower bound are considered more promising to lead to the optimal sensor schedule and thus are expanded first. Sub-trees are pruned if their lower bound is larger than the total cost value of an already completely evaluated sensor schedule.

For a particular node that was reached during the search by employing the sensor schedule $(u_0,\ldots, u_{k-1})$, the total costs can be written as
\begin{equation}
	\label{eq:cost_unkown}
	J(u) = \underbrace{J(u_0,\ldots, u_{k-1})}_\text{known} + \underbrace{J(u_{k}, \ldots, u_{N-1})}_\text{unknown}~,
\end{equation}
where only the value of the first summand is evaluated and thus known. While the value of the second summand is not calculated yet, a lower bound can be easily assigned by exploiting the \SIMs~as shown in the next section. 
Based on this bound, the sub-tree corresponding to the remaining schedules $(u_{k}, \ldots, u_{N-1})$ can be pruned, if the lower bound is larger than a \emph{global bound} $J_\mathrm{min}$, which is the total cost of the currently best completely evaluated sensor schedule. $J_\mathrm{min}$ forms an upper bound of the optimal cost. 
%
%
Obviously, the closer the lower bound to the true value of $J(u_{k}, \ldots, u_{N-1})$, the earlier complete sub-trees can be pruned and thus, the better the pruning performance. 

\oc{\input{alg_ibp}}

\section{Bounding Sensor}
\label{sec:bounding}
Due to the cumulative structure of the total costs $J(u)$ with non-negative summands, a simple lower bound can be obtained by setting the second summand in \eqref{eq:cost_unkown} equal to zero. In the following, this simple bound is referred to as \emph{zero bound} (ZB). This bound can be interpreted as expecting a complete reduction of uncertainty for each time step $k, \ldots, N-1$. Such a reduction can merely be achieved by a sensor information matrix with infinite trace or determinant. 
Obviously, such a \SIM~provides the coarsest lower bound possible.

A significantly tighter lower bound can be derived by means of a so-called \emph{bounding sensor} with \SIM~$\bar \M_k$ for which the covariance matrices evolve according to
\begin{equation}
	\label{eq:bndcov}
	\bar \C_{n+1}^x = \A_{n} \( \(\bar \C_n^x\)^{-1} + \bar \M_n \)^{-1} \A_{n}\T + \C_n^w
\end{equation}
for $n=k,\ldots,N-1$ commencing from $\bar \C_k^x = \C_k^x$. For each time step $n$, this special information matrix fulfills
\begin{equation}
	\label{eq:bndsns}
	\bar \M_n \succeq \M_n^{u_n}
\end{equation}
for all sensors $u_n\in\SS$, where $\bar \M_n$ is as close to $\M_n^{u_n}$ as possible. 
According to \Theo{theo:sensorinfo}, for covariance matrices obtained from \eqref{eq:bndcov} holds $\bar \C_{n+1}^x \preceq \C_{n+1}^x(u_n)$ for all time steps $n=k,\ldots,N-1$ and sensor schedules $(u_k,\ldots,u_{N-1})$. 

To prune the tree at an arbitrary time step (or level) $k$, merely the cost for the bounding sensor needs to be evaluated for the time steps $k, \ldots, N-1$ in order to obtain the lower bound. This calculation can be done in linear time. In comparison, calculating the costs of all possible sensor schedules $(u_k,\ldots, u_{N-1})$ requires exponential computation time.

The complete IBP search algorithm is summarized in \Alg{alg:bb}, where as basic structure a depth-first search is applied. Breadth-first search can be utilized as well. The determination of the bounding sensor with information matrix $\bar \M_k$, which is required for calculating the lower bounds in line~7, is described in the following. 

\subsection{Bounding Sensor for Two Sensors}
Since \SIMs~are symmetric and \PSD, they can be graphically interpreted as ellipsoids similar to the covariance ellipsoids that correspond to covariance matrices. In the following, the ellipsoid corresponding to a \SIM~is referred to as \emph{\SIE}. 
Unlike covariance ellipsoids, sensor information ellipsoids have no distinguished position. Hence, it can be assumed that all \SIE s are centered around the origin.

Based on this interpretation, determining the bounding \SIM~$\bar \M_k$ corresponds to the determination of the covering ellipsoid that contains the ellipsoids of all sensor information matrices. This problem is similar to the so-called L\"owner ellipsoid problem (see e.g. \cite{ Yildirim2006}). Thanks to the fact that all \SIE s have the same center, the L\"owner ellipsoid problem can be significantly simplified.

\begin{figure}[tb]
	\centering
	\includegraphics[]{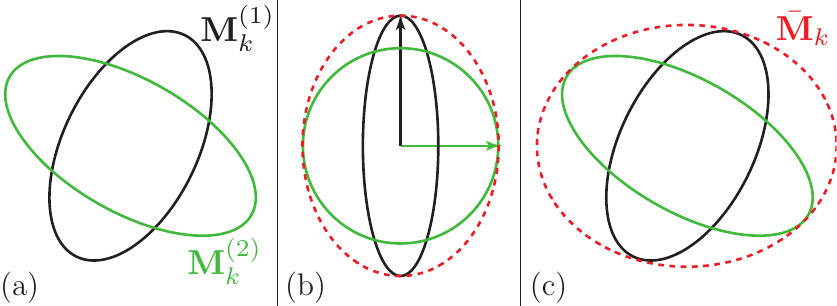}
	\caption{Illustration of the determination of the minimum bounding \SIM~for the two sensors' case. (a) Sensor information ellipsoids of two \SIMs~$\M_k^{(1)}, \M_k^{(2)}$. (b) Result of the simultaneous diagonalization of $\M_k^{(1)}$ and $\M_k^{(2)}$. The covering ellipsoid (red dashed ellipsoid) corresponding to the bounding \SIM~results from taking the maximum eigenvalue (or longest principal axis, indicated by the arrows) for each dimension. (c) Transforming back yields the desired minimum covering ellipsoid corresponding to $\bar \M_k$.}
	\label{fig:covunion}
\end{figure}

At first, the two sensors case is considered. Let $u^{(1)}, u^{(2)} \in \SS$ be the two sensors with information matrices $\M_k^{(1)}, \M_k^{(2)}$\,. For this case, 
a bounding \SIM~with minimum determinant, i.e., a minimum volume covering ellipsoid, results from solving a generalized eigenvalue problem as summarized in the following theorem and as depicted in~\Fig{fig:covunion}.

\begin{theorem}[Min. Bounding Sensor Information Matrix]
	\label{theo:olf_pruning_bndsens}
	The bounding \SIM~$\bar \M_k$ with minimum determinant that fulfills \eqref{eq:bndsns} for two sensor information matrices $\M_k^{(1)}$ and $\M_k^{(2)}$ is given by
	\begin{equation}
		\label{eq:bndinfo}
		\bar \M_k = \(\V\T\)^{-1} \cdot \max\Bigl(\V\T \M_k^{(1)} \V, \V\T \M_k^{(2)} \V\Bigr) \cdot \V^{-1}\, ,
	\end{equation}
	where $\max(\cdot)$ is the element-wise maximum of matrices and $\V$ is the matrix of generalized eigenvectors of $\M_k^{(1)}$ and $\M_k^{(2)}$.
\end{theorem}
\begin{Proof}
At first, the sensor information matrices $\M_k^{(1)}$ and $\M_k^{(2)}$ have to be diagonalized simultaneously, which requires solving the generalized eigenvalue problem
\begin{equation}
	\label{eq:eigen}
	|\M_k^{(1)} - \lambda \M_k^{(2)}| = 0~.
\end{equation}
The solution of \eqref{eq:eigen} is given by the eigenvalues $\lambda_i, i \in \{1, 2, \ldots, n_x\}$ and the matrix of eigenvectors $\V$. This allows diagonalizing $\M_k^{(1)}$ and $\M_k^{(2)}$ according to
\tc{
\begin{align}
	\begin{split}
	\label{eq:diag}
	\V\T \M_k^{(1)} \V &= \diag\([\lambda_1, \ldots, \lambda_{n_x}]\) ~, \\
	\V\T \M_k^{(2)} \V &= \I~.
	\end{split}
\end{align}
}
\oc{
\begin{equation}
	\label{eq:diag}
	\V\T \M_k^{(1)} \V = \diag\([\lambda_1, \ldots, \lambda_{n_x}]\)\quad~,~\quad\V\T \M_k^{(2)} \V = \I~.
\end{equation}
}
Determining the minimum volume covering ellipsoid for diagonal matrices is straightforward by taking the maximum of each diagonal element of the matrices in \eqref{eq:diag} according to
\begin{equation}
	\label{eq:max}
	\max\bigl(\V\T \M_k^{(1)} \V, \V\T \M_k^{(2)} \V\bigr)~.
\end{equation}
This corresponds to taking the longest principal axis for each dimension or equivalently to taking the maximum of $(1, \lambda_i)$ for each $i\in \{1, 2, \ldots, n_x\}$ (see \Fig{fig:covunion}~(b)). Multiplying $\(\V\T\)^{-1}$ from left and $\V^{-1}$ from right to \eqref{eq:max} reverses the diagonalization and leads to the desired minimum \SIM~$\bar \M_k$ (see \Fig{fig:covunion}~(c)).
\end{Proof}

\subsection{Recursive Calculation for an Arbitrary Number of Sensors}
\label{sec:bounding_recursive}
Determining the bounding sensor with minimum \SIM~for an arbitrary number of sensors is computationally expensive in general, as numerical optimization is required. However, a very tight but not necessarily minimum bounding \SIM~can be efficiently calculated by employing the solution of the two sensors' case recursively in a pairwise fashion. 
In doing so, the complexity for determining the bounding \SIM~is linear with the number of sensors. 

The recursion commences from the initial solution $\bar \M^{(2)}_k$ according to \eqref{eq:bndinfo} for the first two sensors $u^{(1)}, u^{(2)} \in \SS$. For the remaining sensors $u^{(i)} \in \SS$, with $i \in \{3, 4, \ldots, |\SS|\}$, the recursion
\begin{equation}
\label{eq:recursive}
	\bar \M_k^{(i)} = \(\V\T\)^{-1} \cdot \max\(\V\T \bar \M_k^{(i-1)} \V, \V\T \M_k^{(i)} \V\) \cdot \V^{-1}
\end{equation}
applies, where $\V$ is the matrix of eigenvectors of $\bar \M_k^{(i-1)}$ and $\M_k^{(i)}$. The final solution $\bar \M_k^{(i)}$ for $i=|\SS|$ is the desired bounding \SIM~$\bar \M_k$ that fulfills \eqref{eq:bndsns}. 


\section{Simulation Example}
\label{sec:sim}
The proposed IBP is evaluated by means of a simplified target tracking example. The state $\rvx_k = [\x_k, \dot{\x}_k, \y_k, \dot{\y}_k]\T$ of the observed target comprises the two-dimensional position $[\x_k, \y_k]\T$ and the velocities $[\dot{\x}_k, \dot{\y}_k]\T$ in $x$ and $y$ direction. The system matrix and noise covariance matrix of $\rvw_k$ of the dynamics model \eqref{eq:problem_system} are 
\begin{equation}
	\label{sec:sim_dynamics}
	\A = \I \otimes \begin{bmatrix} 1 & T \\ 0 & 1\end{bmatrix}
	\ \text{ and }\ 
	\C^w = q \cdot \I \otimes 
	\begin{bmatrix} 
		\tfrac{T^3}{3} &\tfrac{T^2}{2} \\ \tfrac{T^2}{2} & T
	\end{bmatrix}
	~,~\quad
\end{equation}
respectively, where $\otimes$ is the Kronecker matrix product. In \eqref{sec:sim_dynamics},  $T=\Second{1}$ is the sampling interval and $q=0.02$ is the scalar diffusion strength. Mean vector and covariance matrix of the initial state $\rvec x_0$ are $\hat{\vx}_0 = [0, 1, 0, 1]\T$ and $\C_0^x = \I$, respectively. 

A sensor network observes the target. It consists of eight sensors with measurement matrices 
\tc{
\begin{align}
	\H^1 &=
	\begin{bmatrix}
		1 & 0 & 0 & 0
	\end{bmatrix}~,~
	&\H^2 &=
	\begin{bmatrix}
		0 & 1 & 0 & 0
	\end{bmatrix}~,
	\\
	\H^3 &=
	\begin{bmatrix}
		0 & 0 & 1 & 0
	\end{bmatrix}~,~
	&\H^4 &=
	\begin{bmatrix}
		0 & 0 & 0 & 1
	\end{bmatrix}~,\\
	\H^5 &= \begin{bmatrix}(\H_k^1)\T, (\H^3)\T\end{bmatrix}\T~,~ 
	&\H^6 &= \begin{bmatrix}(\H_k^2)\T, (\H^4)\T\end{bmatrix}\T~,\\ 
	\H^7 &= \begin{bmatrix}(\H_k^1)\T, (\H^2)\T\end{bmatrix}\T~,~ 
	&\H^8 &= \begin{bmatrix}(\H_k^3)\T, (\H^4)\T\end{bmatrix}\T~.
\end{align}
}
\oc{
\begin{align}
	\H^1 &=
	\begin{bmatrix}
		1 & 0 & 0 & 0
	\end{bmatrix}~,~
	\H^2 =
	\begin{bmatrix}
		0 & 1 & 0 & 0
	\end{bmatrix}~,~
	\\
	\H^3 &=
	\begin{bmatrix}
		0 & 0 & 1 & 0
	\end{bmatrix}~,~
	\H^4 =
	\begin{bmatrix}
		0 & 0 & 0 & 1
	\end{bmatrix}~,\\
	\H^5 &= \begin{bmatrix}(\H^1)\T, (\H^3)\T\end{bmatrix}\T,~
	\H^6 = \begin{bmatrix}(\H^2)\T, (\H^4)\T\end{bmatrix}\T,~
	\\
	\H^7 &= \begin{bmatrix}(\H^1)\T, (\H^2)\T\end{bmatrix}\T,~ 
	\H^8 = \begin{bmatrix}(\H^3)\T, (\H^4)\T\end{bmatrix}\T.
\end{align}
}
The noise covariance matrices $\C_k^v$ are diagonal, where the diagonal elements are determined randomly from a uniform distribution on the interval $[0,1]$. For each $k$ the weighting matrix $\W_k = \I$, i.e., the objective is to minimize the average estimation error. The cost functions $g_k(\cdot)$ are set to the trace for each $k$. Six different pruning methods are compared for time horizon lengths $N=1,\ldots,8$:\oc{\\[-3ex]}\tc{\\[-1ex]}

\begin{tabular}{rl}
	\GRE & Greedy scheduling, i.e., one-step lookahead.\\
	\CVX & Suboptimal scheduling based on convex \tc{\\ &} optimization \cite{Mo_Automatica2011}.\\
	\ZB & Branch-and-bound with zero bound, see \Sec{sec:bounding}.\\
	\COV & Pruning method proposed in \cite{Logothetis1999}.\\
	\SIP & \ZB~plus the order of sensor information matrices.\\
	\IBP & Proposed information-based pruning.
\end{tabular}
\oc{\\[2ex]}\tc{\\[1ex]}
For each time horizon length, $M = 50$ Monte Carlo runs are performed.

\begin{figure}[tb]
	\centering
	\includegraphics[]{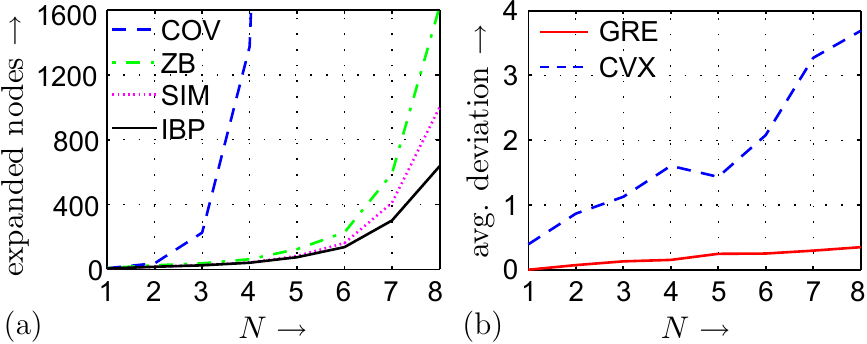}
	\caption{(a) Average number of expanded nodes. (b) Average deviation of greedy scheduling and convex scheduling from optimal scheduling in terms of total costs $J$.}
	\label{fig:sim}
\end{figure}

In \Fig{fig:sim}~(a) the average number of expanded nodes in the search tree is plotted. Compared to the total number of nodes, which is $\sum_{i=0}^N |\SS|^i \in \mathcal O\(|\SS|^N\)$ with $\mathcal O(\cdot)$ being the big-O in Landau notation, optimal pruning significantly reduces the search space. 
The much better performance of \SIP~compared to \ZB~shows the benefit from utilizing the order of \SIMs.
Additionally employing this order can be considered a pre-selection of candidate sensor schedules, while \BB~thins out this candidate set. 
\IBP~provides a much tighter lower bound than \ZB~and \SIP, respectively, which further improves pruning.

The average deviation of the suboptimal methods \GRE~and \CVX~from optimal scheduling is depicted in \Fig{fig:sim}~(b). The average deviation for a given time horizon length $N$ is defined as $\frac{1}{M} \sum_{i=1}^M \(J_{N,\bullet}^i - J_N^i\)$ with $J_N^i$ being the optimal total cost of simulation run $i$ calculated by means of IBP and $J_{N,\bullet}^i$ being the total cost of pruning method $\bullet \in \{\text{\GRE}, \text{\CVX}\}$. 
More significant deviations from the optimal scheduling especially in case of \GRE~are expected for example in scenarios where additional sensor constraints like energy or communication costs (see \Sec{sec:extensions_constraints}) have to be considered, where sensors are temporarily unavailable, or where the dynamics and sensor models are nonlinear (consider for example the results in \cite{Chhetri_EURASIP2006, Xiao_ICASSP2006}). Regarding the runtime, \IBP~is outperformed only by \GRE, while all optimal scheduling algorithms and \CVX~are up to three orders of magnitude slower than \IBP.

\section{Potential Extensions}
\label{sec:extensions}
The proposed information-based pruning algorithm can be extended in many ways in order to apply it to more general sensor scheduling tasks and to further improve the pruning performance. Three possible extensions are discussed.

\subsection{Multiple Sensors per Time Step}
\label{sec:extensions_multiple}
So far, merely one sensor per time step is selected for performing a measurement. 
The extension to multiple sensors requires to schedule a subset of sensors $\mathcal S' \subset \mathcal S$ with $|\mathcal S'| = S'$ per time step. For this purpose, each subset out of ${S \choose S'}$ possible subsets has to be evaluated at each time step if no pruning is performed. 
This additional subset selection has to be included into the tree search in order to employ IBP for reducing the search effort.

A straightforward realization is to perform nested branches per time step instead of a single branch as in the single sensor case. Each nested branch can be considered a tree of depth $S'$ but with a branching factor that decreases with each level. In the first level, the branching factor is $S$, in the second level it is $S-1$ since the sensor selected in the previous branch must not be considered in the current branch. Finally, in level $S'$ the branching factor is reduced to $S-S'+1$. Each path from the root to a leaf node in this subset selection tree corresponds to a sensor subset that can be scheduled for a time step. At each leaf node, the subset selection tree for the next time step begins. Overall the whole search tree has a depth of $N\cdot S'$ with varying branching factor. For comparison, the search tree for the single sensor case has a depth of $N$ with a constant branching factor of $S$. IBP can now be directly employed as in the single sensor case.

\subsection{Sensor Constraints}
\label{sec:extensions_constraints}
In many sensor scheduling problems, there is an additional budget constraint of the form
\begin{equation}
	\mat B \cdot \vec u \le \vec b~,
\label{eq:extensions_constraint}
\end{equation}
where $\vec u = [\vec u_0\T, \ldots, \vec u_{N-1}\T]\T \in \{0,1\}^{N\cdot S}$ is a binary sensor scheduling vector with elements $\vec u_k \in  \{0,1\}^S$ for each time step $k=0,\ldots,N-1$. Such a budget can for example be a maximum number of sensors $K \in \NewN$ allowed to perform measurements. Here, $\mat B = [1, 1, \ldots, 1]\T$ is a vector of ones and $b = K$. A further example is an energy budget $K_i \in \NewN$ that is assigned to each sensor $i$. The energy budget $K_i$ is reduced by one after each measurement of sensor $i$. In this case, $\vec b = [K_1, K_2, \ldots, K_S]\T$ and $\mat B$ is a binary matrix, where only the elements $i, i+S,\ldots,i+(N-1)\cdot S$ of row $i$ of $\mat B$ are equal to one.

If such sensor constraints are present, IBP cannot be applied directly. This is basically due to pruning based on the partial ordering of sensors as this ordering does not take the sensor constraints into account. 
Additionally to considering the sensor information matrices, it is also necessary to check whether the budget constraint can still be fulfilled if a node of the search tree is pruned.
Consider the example mentioned above, where a maximum number of sensors $K \in \NewN$ is allowed to perform measurements. To map IBP on this constrained scheduling problem, one possibility is to include a virtual sensor with information matrix $\mat M = \mat 0$. Selecting this virtual sensor corresponds to performing no measurement. No budget costs are assigned to this virtual sensor, i.e., the elements in $\mat B$ corresponding to the virtual sensor are zero. Of course, the information matrix of the virtual sensor is always dominated by all other sensors, but selecting the virtual sensor will never violate the constraint $K$. 
Thus, by ensuring that this sensor is never pruned, 
IBP can be applied directly.

In contrast to the partial ordering, calculating the lower bound based on the \SIMs~is always directly possible. Certainly, the bound will not be as tight as in the unconstrained case. Calculating the bound at a particular node in the search tree requires to assume the best case where each sensor is available at all times. Thus, all sensor information matrices are incorporated into calculating the bounding sensor even if some sensors might not be available in future. 

It is worth mentioning that the constraint \eqref{eq:extensions_constraint} itself can be considered as some kind of pruning. Avoiding the  selection of a sensor that will violate the constraint corresponds to pruning the whole sub-tree of this particular sensor.

\subsection{Improving the Bounds}
\label{sec:extensions_bounds}
Scheduling algorithms based on convex optimization (see for example \cite{Huber_CIP2010, Joshi_TSP2009, Mo_Automatica2011}) are sub-optimal, but they can be employed in an optimal \BB~algorithm for calculating lower bounds as shown in \cite{Huber_CIP2010}. In many cases, these bounds are tighter than the bound proposed in this paper. However, calculating lower bounds based on convex optimization is computationally significantly more complex. 
Thus, both bounds could be combined, where the proposed bound is used firstly for pruning in order to reduce the need of calculating bounds via convex optimization.

For IBP merely the total cost of the currently best completely evaluated sensor schedule is used for setting the global bound value $J_\mathrm{min}$. By means of tight upper bounds, $J_\mathrm{min}$ can be decreased more rapidly, which will significantly improve the pruning performance. One way to calculate upper bounds is to use Kalman predictions without measurement updates. Better upper bounds can be provided, if the \SIMs~are used for calculating a further bounding sensor with information matrix $\M_k' \preceq \M_k^{u_k}$ similar to \eqref{eq:bndsns}. Therefore, the recursive algorithm proposed in \Sec{sec:bounding} can be employed, where $\max$ in \eqref{eq:max} and \eqref{eq:recursive} is replaced by~$\min$\,.

\section{Conclusion}
\label{sec:discussion}
The proposed information-based pruning excludes complete sub-trees very early, while preserving the optimal sensor schedule is guaranteed. Compared to existing optimal pruning methods, the effectiveness of the proposed method relies on exploiting the properties of the \SIMs. This allows pruning without explicitly evaluating the Riccati equation and by calculating a tight lower bound to the total costs. \IBP~can be extended for example to schedule multiple sensors per time step or to consider constraints like energy budgets.

\section{Acknowledgments}
\label{sec:ack}
The author gratefully thanks Yilin Mo and Bruno Sinopoli from the Department of Electrical and Computer Engineering, Carnegie Mellon University, Pittsburgh, USA, for providing the source code of their algorithm proposed in \cite{Mo_Automatica2011}. This research work was partially supported by the Intelligent Sensor-Actuator-Systems Laboratory (ISAS), Karlsruhe Institute of Technology (KIT), Germany, as well as by the Fraunhofer Institute of Optronics, System Technologies and Image Exploitation IOSB, Karlsruhe, Germany.

\bibliographystyle{IEEEtran}

\end{document}